\documentclass[letterpaper]{article} % DO NOT CHANGE THIS

\usepackage{aaai24}  % DO NOT CHANGE THIS
\usepackage{times}  % DO NOT CHANGE THIS
\usepackage{helvet}  % DO NOT CHANGE THIS
\usepackage{courier}  % DO NOT CHANGE THIS
\usepackage[hyphens]{url}  % DO NOT CHANGE THIS
\usepackage{graphicx} % DO NOT CHANGE THIS
\def\UrlFont{\rm}  % DO NOT CHANGE THIS
\usepackage{natbib}  % DO NOT CHANGE THIS AND DO NOT ADD ANY OPTIONS TO IT
\usepackage{caption} % DO NOT CHANGE THIS AND DO NOT ADD ANY OPTIONS TO IT
\usepackage{algorithm}
\usepackage{algorithmic}

\usepackage{booktabs}
\usepackage{multirow}
\usepackage{color}
\usepackage{comment}

\usepackage{color}

\usepackage{soul}
\usepackage{listings}

\usepackage{booktabs}
\usepackage{multirow}

\definecolor{mGreen}{rgb}{0,0.6,0}
\definecolor{mGray}{rgb}{0.5,0.5,0.5}
\definecolor{mPurple}{rgb}{0.58,0,0.82}

\lstdefinestyle{CStyle}{
    backgroundcolor=\color{white},   
    commentstyle=\color{mGreen},
    keywordstyle=\color{blue},
    numberstyle=\tiny\color{mGray},
    stringstyle=\color{mPurple},
    basicstyle=\footnotesize,
    frame=single,
    breakatwhitespace=false,         
    breaklines=true,                 
    captionpos=b,                    
    keepspaces=true,                 
    numbers=left,                    
    numbersep=5pt,                  
    showspaces=false,                
    showstringspaces=false,
    showtabs=false,                  
    tabsize=2,
    language=C
}

\usepackage{newfloat}
\usepackage{listings}
\DeclareCaptionStyle{ruled}{labelfont=normalfont,labelsep=colon,strut=off} % DO NOT CHANGE THIS
\floatstyle{ruled}
\newfloat{listing}{tb}{lst}{}
\floatname{listing}{Listing}
\title{Code Security Vulnerability Repair Using Reinforcement Learning \\with Large Language Models}
\author {
    Nafis Tanveer Islam,\\
    Mohammad Bahrami Karkevandi, \\
    Peyman Najafirad \footnote{Corresponding Author}
}
\affiliations {
    Secure AI an Autonomy Laboratory\\
    Department of Computer Science \\
    University of Texas at San Antonio\\
}
\begin{document}

\maketitle

\begin{abstract}
With the recent advancement of Large Language Models (LLMs), generating functionally correct code has become less complicated for a wide array of developers. While using LLMs has sped up the functional development process, it poses a heavy risk to code security. Code generation with proper security measures using LLM is a significantly more challenging task than functional code generation. Security measures may include adding a pair of lines of code with the original code, consisting of null pointer checking or prepared statements for SQL injection prevention. Currently, available code repair LLMs generate code repair by supervised fine-tuning, where the model looks at cross-entropy loss. However, the original and repaired codes are mostly similar in functionality and syntactically, except for a few (1-2) lines, which act as security measures. This imbalance between the lines needed for security measures and the functional code enforces the supervised fine-tuned model to prioritize generating functional code without adding proper security measures, which also benefits the model by resulting in minimal loss. Therefore, in this work, for security hardening and strengthening of generated code from LLMs, we propose a reinforcement learning-based method for program-specific repair with an attempt to combine semantic and syntactic reward mechanisms that focus heavily on adding security and functional measures in the code, respectively.
\end{abstract}

\section{Introduction}
\label{1_Introduction}

%Software vulnerabilities are security flaws, glitches, or weaknesses found in software systems that could be exploited by attackers to undertake malicious activities [6]. In particular, criminal groups may make use of unresolved security vulnerabilities in software to attack and damage a system to steal confidential information or extort assets, resulting in severe economic damage [17]. According to the statistics of the National Vulnerability Database (NVD), the number of vulnerabilities discovered per year is considerably increased five times from 4k+/year in 2011 to 20k+/year in 2021 [1].

%Software vulnerabilities, such as buffer overflows and SQL injections, have a critical impact on global economies and can harm millions of users. Once a vulnerability is discovered, it is often crucial to fix it promptly to minimize the potential for exploitation.

% Paragraph 1
A software vulnerability is a possible set of flaws or weaknesses in the system that allows an attacker to gain access to the system, halt the service provided by the system, or ask for ransom from the software vendors. This becomes even more challenging when threat actors use malicious techniques to steal confidential information or cause economic damage \cite{dowd2006art}. In an era of escalating cyber threats, the Open Source Software (OSS) community stands at an unprecedented crossroads, safeguarding the foundation of countless software systems and products integral to modern society. Nevertheless, security vulnerabilities in poorly written software play a vital role across government and its infrastructure, causing widespread disruption of services \cite{synopsysrisc}.

\begin{figure*}[h]
    \centering
    \includegraphics[scale=0.55]{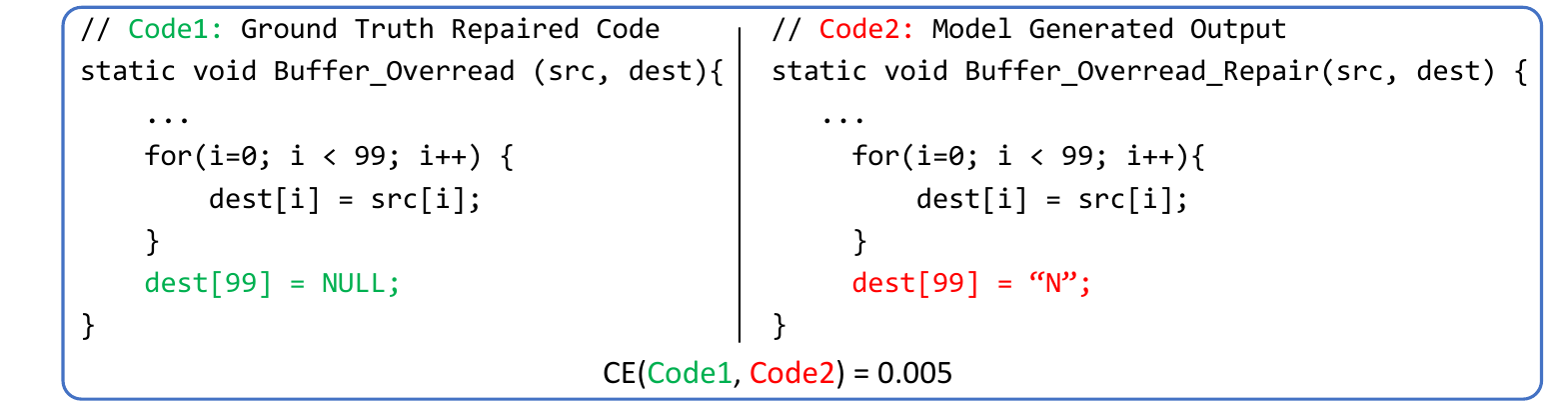}
    %\vspace{-0.3cm}
    \caption{An illustrative example showcases how the Cross-Entropy (CE) emphasizes only functionality and neglects security. \color{black}}
    \label{fig:fig_1}
    %\vspace{-3mm}
\end{figure*}

The wide use of large language models further enhances the security vulnerability. Their functional correctness is heavily challenged by the potential introduction of security issues within the generated code \cite{sandoval2023lost, pearce2022examining}. Despite the democratization of coding, providing increased accessibility and productivity among developers from code generated by large language models frequently falls short of established software security standards, possibly harboring vulnerabilities in approximately 40\% of cases \cite{pearce2022asleep}. This security challenge extends beyond individual models, with various evaluations revealing that other cutting-edge LMs \cite{nijkamp2022codegen, fried2022incoder}, similar to Copilot, display identical security issues, as highlighted in \cite{li2023starcoder}. A separate investigation carried out by \cite{khoury2023secure} identified that ChatGPT, in 16 out of 21 security-relevant cases, generates code falling below minimal security standards. The more alarming concern is the training data of generative AI models, deeply rooted in the same insecure public repositories \cite{chen2021evaluating}. As a result, unless pre-trained explicitly in security, generative AI models give rise to a vicious cycle of generating functional code by disseminating vulnerability.

%The landscape grows more intricate with emerging technologies like ChatGPT, which introduces significant security challenges despite enhancing developers' coding ability. \cite{khoury2023secure, sandoval2022security}. Open-source code is highly accessible, so novice developers are likely to use it to complete the functionality of their coding task. Although this democratization of coding has increased the developers' productivity by enabling more people to engage in programming \cite{democr1, democr2}, open-source code available online or produced by large language models often fails to meet software security standards, potentially containing vulnerabilities in around 40\% of cases. \cite{pearce2022asleep}. More concerning is the training data of generative AI models, deeply rooted in the same insecure public repositories. As a result, unless trained explicitly with security in mind, generative AI models (for automated code generation) give rise to a vicious cycle of generating and disseminating more insecure code. Since the use of external code persists from novice to experienced developers, the possibility of spreading vulnerable code across various applications remains high.

%% ***** START WORKING ON THIS PARAGRAPH ******** %%
The pre-training of large language models generally follows the pre-training procedure for generating texts. CodeT5 \cite{wang2021codet5} used three pre-training methods to pre-train the model on code. Initially, they tasked the model with predicting a masked set of tokens of code; then, they tasked the model with tagging the identifiers of the code tokens and, finally, predicting the identifiers. StarCoder \cite{li2023starcoder} pre-trains their model by training it with masked language modeling and next sequence prediction. For functional code generation of LLMs, CodeRL \cite{le2022coderl} fine-tunes the model to degenerate programs based on feedback from example unit tests and critic scores.

Furthermore, functional code generation is a task where the model has to generate entire code, which is based on specific instructions. The functional task of Code 1 and Code 2 is to copy data from the source to the destination. However, Code1 and Code 2 in Figure \ref{fig:fig_1} demonstrate that security repair merely adds a few lines to the code. Here, Code 1 is the ground truth repaired function, and Code 2 is the model-generated outcome that attempts to repair a vulnerable function. The original vulnerable input to the model is the exclusion of the line \texttt{dest[99] = NULL}. From the ground truth. We observe that the cross-entropy (CE) loss between the original and model outcomes is very low. Thus, the model is tempted to ignore the security measures and generate code almost similar to the input since it yields minimum loss. This issue effectively demonstrates that the use of CE loss only provides us with token-based similarity when we compare the generated code with the ground truth. However, this also demonstrates that merely checking code similarity without checking the semantics and syntactic does not do a proper evaluation of the code. 

%  Moreover, BERTScore on Code 1 and Code 2 shows slightly improved loss values. Similarly, if we only compare the security measure with a syntactic similarity-based measurement CodeBLEU, we still see that the loss value remains much higher. These higher values indicate that cross-entropy loss is insufficient to make the generative model security aware. 

%Furthermore, code with similar vulnerabilities needs a similar solution based on the functionality of the specific code.

To address these challenges of generating secure code while maintaining functionality, we used a reinforcement learning (RL) based technique for vulnerability repair of the source with a specialized reward function designed specifically for enhancing code security. To address the issue of security-focused comparison of the generated code and ground truth, we propose a syntactic reward mechanism that rewards the RL model only when proper security measures are added to the code. Furthermore, to ensure the functional correctness of our proposed system, we propose a syntactic and semantic reward function. Therefore, we propose \textit{SecureCode} powered by a large language model CodeGen2 \cite{nijkamp2023codegen2} trained using reinforcement learning with the combination of syntactic and semantic reward to generate security repairs of vulnerable code that are functionally correct. 

%To address the critical need for identifying code vulnerability code and provide repairs to developers on vulnerable code, we are the first to introduce a reliable AI-assisted solution \texttt{SecRepair} for vulnerability repair, using reinforcement learning. Our system leverages the power of CodeGen2 \cite{nijkamp2023codegen2}, a large language model designed and explicitly fine-tuned for code security analysis to identify and repair vulnerable code.

In summary, the contributions of this paper are:
\begin{itemize}
    \item We introduce a state-of-the-art LLM-powered tool to repair code vulnerabilities written in C/C++. Our proposed system leverages reinforcement learning to generate secure code while maintaining functionality.
    
    %Our system leverages Reinforcement Learning with syntactic and semantic reward values to enhance its capabilities of generating secure and functional code. 

    \item We introduce the combination of syntactic and semantic reward values to enhance the capability of our proposed reinforcement learning model to generate secure and functional code.
    %The dataset contains detailed instructions for localizing, repairing, and describing vulnerabilities, facilitating precise analysis and remediation. 

    \item Our research also includes a quantitative analysis of the results to demonstrate the security repair quality of our proposed model.

\end{itemize}

\section{Related Works}
\label{2_related_work}

Code vulnerabilities have profound implications across diverse domains in the digital realm, ranging from the utilization of digital devices within IoT ecosystems and online accounts \cite{atashpanjeh2022intermediate} to pivotal systems like containers \cite{haq2022security} and operating systems. Although anticipating specific sophisticated techniques proves challenging, most of these vulnerabilities can be attributed to developers' setbacks in ensuring robust code security. This intersection of AI and cybersecurity emphasizes the critical role of proactive measures in fortifying our digital infrastructure against potential threats.

\paragraph{\textbf{Code Vulnerability with LLMs:}}
Most Large Language Models (LLMs) exhibit minimal concern for security aspects in programming languages. Pearce \cite{pearce2022examining} scrutinized LLMs' zero-shot vulnerability repair capabilities and the challenges posed by bugs. Experimental findings indicated that while LLMs can generate bug fixes, they require a specifically crafted prompt for addressing a particular bug. SVEN \cite{he2023controlling} introduced an adversarial technique to assess LLMs, proposing generating safer code by utilizing property-specific continuous vectors to guide program generation. Additionally, studies by \cite{pearce2022asleep}, \cite{jesse2023large}, and \cite{sandoval2022security} extensively explored autocompletion effectiveness, integrating LLMs with various IDEs and analyzing outcomes. While \cite{pearce2022asleep} and \cite{jesse2023large} concluded that LLMs produce vulnerable code, Sandoval \cite{sandoval2022security} suggested that LLMs contribute to generating more functional code with improved security features.

\paragraph{\textbf{Vulnerability Repair:}} Repairing programs pose a formidable challenge, requiring the identification of vulnerable lines and the subsequent generation of a suitable compilable line for fixing the vulnerability. Earlier approaches, such as those incorporating human-specified safety properties \cite{huang2019using}, enforced constraints like preventing access to memory beyond program boundaries. Zhang et al. \cite{zhang2022program} introduced methods to identify patch invariants or vulnerable code locations, employing a set of templates for repair generation. More recent approaches like Vrepair \cite{chen2022neural} adopt a transformer-based transfer learning strategy to address vulnerabilities in real-world programs. Similarly, VulRepair \cite{fu2022vulrepair} presents a vulnerability repair technique utilizing the pre-trained CodeT5 \cite{wang2021codet5} and BPE tokenizer. However, with the emergence of code-based Large Language Models (LLMs) like Codex \cite{chen2021evaluating}, works by Pearce et al. \cite{pearce2022examining}, Jesse et al. \cite{jesse2023large}, and Prenner \cite{prenner2021automatic} demonstrate some capability of these models to repair vulnerable code through zero-shot learning.

% \textcolor{red}{Adding this paragraph and providing some examples of why poeple used RL for LLMs could address the concerns of reviewers that why Supervised Fine-Tuning is not as much effective.} 

\paragraph{\textbf{Fine-Tuning of LLMs with RL:}} Despite the simplicity of fine-tuning a neural network using Supervised Fine-Tuning, recent advancements in Large Language Model training \cite{touvron2023llama, ouyang2022training} show that although Reinforcement Learning is less intuitive for the human mind, it is highly effective in guiding Large Language Models to follow specific requirements, in our case, security of the code. Additionally, Reinforcement Learning provides a framework to use arbitrary weak signals to align the model. These signals include human preference and quantitative evaluation scores such as BLEU \cite{papineni2002bleu}, which may not be possible or computationally expensive to utilize in a supervised learning setting due to their non-differentiable nature \cite{Liu2019OptimizingBS}. One way to overcome this limitation for Supervised Fine-Tuning is to use differentiable loss functions such as the Cross-Entropy Loss. However, our initial analysis in Figure \ref{fig:fig_1} shows that Cross-Entropy Loss does not perform well in a code security measurement setting. Also, Since SFT signals are strong and explicit, a model fine-tuned using supervised settings may be more prone to losing generality \cite{lin2023speciality}. Aggregating these reasons motivates more research on using reinforcement learning to generate secure code.

\section{Methodology}
\label{3_methodology}

%Sequential Context Understanding: Causal decoders excel in understanding and generating sequences of data. In code-related tasks, understanding the sequential nature of statements and their dependencies is crucial, and causal decoders inherently capture this.

%Long-range Dependencies: Causal models are designed to handle long-range dependencies in sequences. In code, where the context of a variable or function call might span several lines or even pages, the ability to capture these dependencies is vital for accurate code generation or repair.

%Decoder models use only the decoder of a Transformer model. At each stage, for a given word the attention layers can only access the words positioned before it in the sentence. These models are often called auto-regressive models.

\begin{figure*}[t]
    \centering
    \includegraphics[scale=0.60]{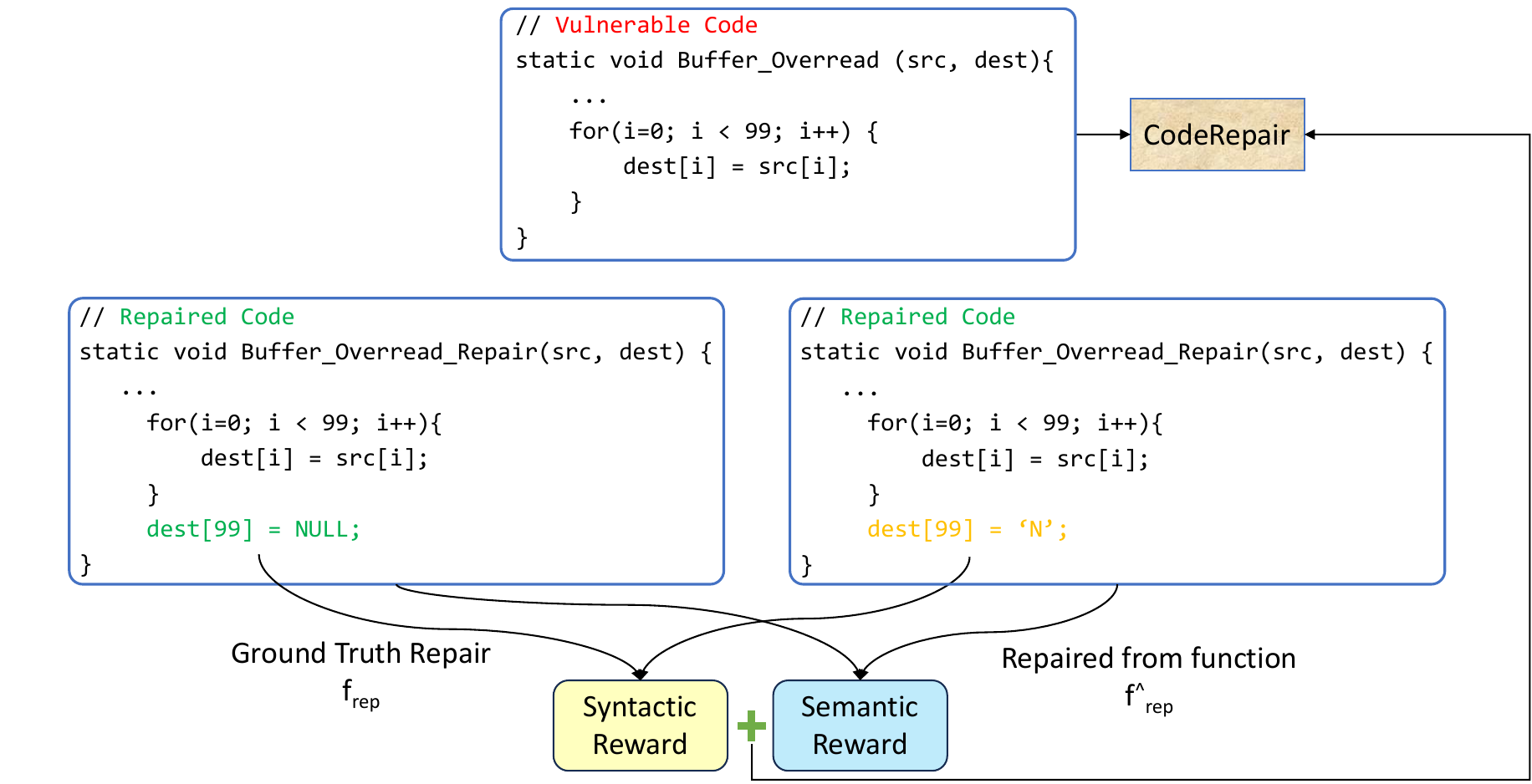}
    %\vspace{-0.3cm}
    \caption{A high-level overview of our proposed CodeRepair System with Reinforcement learning with semantic and syntactic loss.}
    \label{fig:fig_2}
    %\vspace{-3mm}
\end{figure*}

One of the major challenges in code generative models is the long-range dependency of source code \cite{islam2023unbiased}, where the declaration and the actual usage of a variable are far apart. Because of the long-range dependency, it is challenging for an LLM to determine the position where to add security measures for a given code. Therefore, we use a causal-decoder model, CodeGen2, that emphasizes the importance of previous tokens for our code repair work with enhanced generalization capability \cite{wang2022language}. In code, the context of a variable or function call might span several lines or even pages; the ability to capture these dependencies is critical for accurate code generation or repair, which a causal decoder model can use effectively and efficiently. Regular encoder-decoder-based LLM architecture shows a bidirectional property where they read tokens from forward and backward. Such a bidirectional architecture causes unnecessary computation that is not needed for code vulnerability repair since the reasoning of a vulnerability always exists before the code breaks or has runtime issues. Moreover, causal decoder models remove the encoder layer. Therefore, they have faster inference time with less memory footprint. Figure \ref{fig:fig_2} demonstrates the overall architecture of our proposed system.

%One of the major challenges with code vulnerability repair is that the code sample can be composed of a high number of tokens. Moreover, code writing, execution, and repair is a sequential task that requires a sequential understanding of the code tokens. However, regular encoder-decoder-based LLM architecture shows a bidirectional property where they read tokens from forward and backward. Such a bidirectional architecture causes unnecessary computation that is not needed for code vulnerability repair. Therefore, we introduce a simple but effective modification to encoder-decoder architecture for longer code sequences by dropping the encoder module. Then we combine the input and output sequences into a single sequence and train as a standard language model. In addition to faster inference time, we also decrease the memory footprint by almost half. 

Initially, we convert the vulnerable input function $f_{vul}$ and the ground truth repaired output function $f_{rep}$ into a sequence of tokens, $t_1, t_2, ... t_p \in T$ and $y_1, y_2, ... y_q \in Y$. Then we combine both into a unique sequence $w_1, w_2, ... w_{p+q}$ = $(t_1, ..., t_p, \$,  y_1, ..., y_q)$, where the $\$$ is a special token separating the vulnerable and the repaired program. $p$ is the number of input tokens, and $q$ is the number of output tokens. 

% We train the model with  auto-regressively to predict the next word given the previous ones using the following equation:
% \begin{equation}
%     p(w_1, w_2, ... w_{p+q}) = \Pi_{j=1}^{p+q}   p(w_i|w_1, ..., w_{j-1})
% \end{equation}

\subsection{Reinforcement Learning for Functional Repair}

%After generating a proper developer-friendly description with code repair, at stage 3 of Figure \ref{fig:fig_2}, we clone our original pre-trained SecRepair and employ it to generate code comments for reinforcement learning. 

Given the causal decoder architecture of \texttt{SecureCode}, our model is forced to predict the next code token, and the model cannot overlook future tokens by looking at the next token during output generation. The model is provided with the input sequence $t_i \in T$  during inference, which auto-regressively generates the output, $y_i$.

Therefore, we fine-tune our model with reinforcement learning for a code-to-code generation task where the generated code is the repaired version of the original input code. To achieve functional code repair, our proposed reinforcement learning technique combines syntactically and semantically aware reward functions with the Proximal Policy Optimization (PPO) algorithm proposed by Schulman et al. \cite{schulman2017proximal}. We denote the input function or the vulnerable code as $f_{vul}$. The repaired output generated the model as $\hat{f_{rep}}$, and the ground truth repaired function is $f_{rep}$. We assume the output sequence is $w^r_1, w^r_2, ... w^r_k$, where $k$ is the total number of output tokens of the generated repair.

We define the token-wise code generative process as a deterministic Contextual Markov Decision Process \cite{hallak2015contextual}  with observable context only from previous tokens of vulnerable code. The repaired code sequence generated, which is the state at the $k^{th}$ token generation\textcolor{black}{,} is defined by our policy $\pi (. | w^r:k-1, t)$, which is the probability distribution of the previous \textcolor{black}{$k-1^{th}$ input tokens from vulnerable function $f_{vul}$}.

%- I think what making the readers confused of whether the policy is word by word or sequnce by sequnce is that we can not say "$k-1^{th}$"(singular) input "tokens"(plural). 

\paragraph{Policy Optimization:} The reinforcement learning objective is to find the optimal policy by maximizing the cumulative syntactic and semantic reward by adding security measures while keeping code functionality checked.

To meet this requirement, we introduce a syntactic code evaluation technique, CodeBLEU \cite{ren2020codebleu}, as a reward value to check the syntactic similarity between the repaired lines generated by the model and the ground truth repair. Here in Figure 2, the repaired line refers to the new line the model is expected to generate to add security measures to the code. Furthermore, to ensure the functional correctness in the generated outcome, we utilize BERTScore \cite{zhang2019bertscore} as a semantic reward value that quantifies the semantic similarity or functionality between the entire input vulnerable code and the generated repaired code. If the original vulnerable code is $f_{vul}$, the repaired code generated by the model is $\hat{f_{rep}}$, and the ground truth repaired code is $f_{rep}$. As such, we calculate the policy optimization $r_\theta$, using the following equation:
\begin{equation}
    \label{eqn:3}
    L(r_\theta) =  log (\sigma(r_\theta (f_{vul}, f_{rep}) - r_\theta (f_{vul}, \hat{f_{rep}}) )
\end{equation}

where $r_\theta (f_{vul}, f_{rep})$ and $r_\theta (f_{vul}, \hat{f_{rep}})$ is the scalar output of the reward model for the vulnerable code  $f_{vul}$. Here $\sigma$ is an activation function, and $\theta$ is a learnable parameter.

%BERTScore focuses on computing semantic similarity between tokens of reference and hypotheses rather than computing token-level syntactical similarity. We use BERTScore to produce a reward value that ranges from 0.0 to 1.0, where 0.0 implies the lowest reward, and 1.0 implies the highest reward.

%E_{D\_c, \hat{D\_c}}

\paragraph{Reward:} We calculate the reward by combining the CodeBLEU \cite{ren2020codebleu} score and BERTScore \cite{zhang2019bertscore}. CodeBLEU is the weighted combination BLEU score \cite{papineni2002bleu}, $BLEU_{weight}$ is the weighted n-gram match, obtained by comparing the generated code and the ground truth repaired code tokens, $Match_{ast}$ is the AST match, exploring the syntactic information of code, and $Match_{df}$ is the dataflow match, considering the similarity between ground truth and generate code.

\begin{equation}
\label{eqn1}
    R_{CodeB} = \alpha . B + \beta . B_{weight} + \\ \delta . Match_{ast} + \gamma . Match_{df}
\end{equation}

Here, $B$ in the equation stands for BLEU.

Furthermore, we use BERTScore for semantic comparison using cosine similarity score. A BERT vector represents tokens that permit the generation of a soft similarity measure instead of exact matching. The cosine similarity of a reference token from ground truth repair $t_r^i$ and a candidate token $\hat{w_r^i}$, we calculate the cosine similarity as $(t_r^i)^T \hat{w_r^i}$. Therefore, the F1 measurement of the BERTScore stands as follows; 

\begin{equation}
\label{eqn2}
    R_{BERT} = \frac{1}{|t_r^i|} \sum_{t_r^i \in T} max (t_r^i)^T \hat{w_r^i}
\end{equation}

%\textcolor{red}{Clarify what are i and r. especially $t_r^i \in T$ is undefined. ts in T only had one subscript}

where $R_{BERT}$ is our expected BERTScore. We combine Equation \ref{eqn1} and \ref{eqn2} to get the final reward value. The final reward is calculated as follows:

\begin{equation}
    R = R_{CodeB} + R_{BERT}
\end{equation}

Here, $R$ is the final calculated reward value.

\section{Experiments}
\label{4_experiments}

\subsection{Dataset}
We use the VulDeeLocator dataset \cite{li2021vuldeelocator} for C/C++ programs with vulnerable source code and their corresponding repairs. The code is sourced from the National Vulnerability Database (NVD) and the Software Assurance Reference Dataset (SARD). This dataset has 40,382 vulnerable and 115,157 not-vulnerable code snippets.

\subsection{Evaluation Metrics}

\paragraph{\textbf{BLEU Score:}}
The BLEU \cite{papineni2002bleu} score evaluates machine-generated text where the score ranges between 0 and 1. A value of 0 means that the generated output does not have a single n-gram match with the original output, while 1 means a perfect n-gram match with the original outcome. However, it's important to note that BLEU is a reference-based metric and may not capture all aspects of translation quality, such as fluency or semantic accuracy. The final BLEU score is calculated using sentence-level n-gram precision and the brevity penalty. The n-gram precision is weighted based on the length of the n-grams (from unigrams to a specified maximum n-gram order, typically 4). The geometric mean of the weighted n-gram precisions is multiplied by the brevity penalty to obtain the BLEU score.

\paragraph{\textbf{Rouge-L:}} Similar to BLEU, Rouge-L \cite{lin2004rouge} score is also a number between 0 and 1 to measure the similarity of two generated texts. It generates a score by quantifying precision and recall by examining the longest common subsequence (LCS) between the generated and reference codes. Precision, in this context, refers to the ratio between the length of the LCS and the length of the generated code, whereas recall refers to the ratio between the length of the LCS and the length of the reference code. These metrics reveal the degree of consistency between the generated code and the reference code, highlighting their similarity and concordance. ROUGE-L is a valuable metric for evaluating the effectiveness of code vulnerability analysis algorithms by quantifying the quality of the generated code relative to the reference code.

% \usepackage{booktabs}
% \usepackage{multirow}
\begin{table}[t]
\centering
\begin{tabular}{@{}llll@{}}
\toprule
\textbf{Model}                                                              & \textbf{Parameter} & \textbf{BLEU} & \textbf{Rouge-L} \\ \midrule

\multirow{3}{*}{\begin{tabular}[c]{@{}l@{}}CodeGen2 \\ (SFT) \end{tabular}} & 1B                 & 0.50          & 0.56                         \\
& 3.7B               & 0.70          & 0.69                       \\
& 7B                 & 0.80          & 0.81                         \\ 
\midrule

\multirow{3}{*}{\begin{tabular}[c]{@{}l@{}}CodeRepair \\ (RL) \end{tabular}} & 1B                 & 0.70          & 0.73                         \\
& 3.7B               & 0.76          & 0.80                          \\
& 7B                 & \textbf{0.86}          & \textbf{0.88}                         \\ \bottomrule
\end{tabular}
\caption{BLEU, Rouge-L Evaluation on generating code repairs from vulnerable code inputs.}
\label{tab:tab1}
\end{table}

\subsection{Experimental Results and Discussion}

In our experiments, we split datasets randomly into 80:10:10 for training, validation, and testing. We utilized a pretrained CodeGen2 model with 32 decoder layers. We trained for 20 epochs with a max token length of 512. Our 7B parameter model uses a learning rate of $2e^{-5}$, a batch size of 2, a beam size of 4, and a temperature value of 0.5 for optimal performance. The training was conducted on 8 NVIDIA A100 GPUs with 40 GB of memory each. Integration of DeepSpeed \cite{rajbhandari2020zero} with our HuggingFace CodeGen2 implementation for minimized memory footprints.

To ensure our proposed system is repairing vulnerability accurately and reliably, we compare our proposed method with the regular fine-tuning-based method. We did two types of training to generate repairs of the vulnerable code. We used 1B, 3.7B, and 7B variants of the CodeGen2 model for comparison. We trained these three variants using Supervised Fine Tuning (SFT) and our proposed Reinforcement Learning (RL) technique for comparative analysis. In order to quantitatively evaluate the repairs, we use BLEU and Rouge-L metrics.

\paragraph{Discussion:}
Our empirical results highlight the superior performance of our model when compared to counterparts with comparable or fewer parameters. As outlined in Table \ref{tab:tab1}, our model exhibits a notable improvement, surpassing the 7B variant of CodeGen2 (SFT) by 0.06 BLEU and 0.07 Rouge-L score. Particularly noteworthy is the efficacy of our proposed Reinforcement Learning (RL) training approach, outperforming the SFT method across the other two variants. A consistent trend emerges in the consistently higher Rouge-L scores, underscoring its robustness in evaluating similarity based on the longest common subsequence (LCS) between original and generated tokens, as opposed to BLEU's n-gram-centric methodology.

%Our experimental findings underscore the superiority of our model over counterparts with similar or fewer parameters. In Table \ref{tab:tab1}, our model surpasses the 7B variant of CodeGen2 (SFT) by 0.06 BLEU and 0.07 Rouge-L score. Notably, our proposed RL training method outperforms the SFT method for the other two variants. A noteworthy trend is the consistently higher Rouge-L scores, emphasizing its reliability in evaluating similarity based on the longest common subsequence (LCS) between original and generated tokens, in contrast to BLEU's n-gram-based approach.

%Our experimental results demonstrate the superiority of our model compared to other models with similar or fewer parameters. In Table \ref{tab:tab1}, we observe that the model achieves a 0.06  BLEU higher than the 7B variant of CodeGen2 (SFT) and 0.07 Rouge-L score. Furthermore, we can see that for the other two variants, our proposed RL method of training the model achieves superior performance compared to the SFT method. An intriguing observation from the comparison of BLEU and Rouge-L scores is that, in most instances, the Rouge-L score remains consistently higher. This phenomenon arises because the BLEU score evaluates similarity based on n-grams and compares the match between those n-grams. In contrast, the Rouge-L score accurately measures similarity using the least common sequence (LCS) between the original and generated output tokens, providing a more reliable evaluation of the generated code.

\vspace{4mm}
\begin{lstlisting}[style=CStyle, label={lst:lst1}, caption=Case Study 1: Vulnerability Repair in Mutex Condition of a Function]
void RecordMutation(TF_Graph* graph, const TF_Operation& op,
                    const char* mutation_type) {
  // If any session has already run this node_id, mark this session as
  // unrunnable.
  for (auto it : graph->sessions) {
    {
      mutex_lock session_lock(it.first->mu);
      if (it.first->last_num_graph_nodes > op.node.id()) {
        it.second = strings::StrCat(
            "Operation '", op.node.DebugString(), "' was changed by ",
            mutation_type,
            " after it was run by a session. This mutation will have no effect, "
            );
      }
    }
    // Releasing Lock -- Repaired Line
    mutex_unlock session_unlock(it.first->mu);
  }
}
\end{lstlisting}
\vspace{4mm}

\section{Case Study of the Generated Outcomes}

In this section, we will provide three case studies which will compare the outcome of our RL model vs. the SFT model.

\paragraph{Case Study 1}In this section, we will analyze the qualitative outcome between the SFT- and RL-based fine-tuning models. The initial implementation of the complex algorithm exhibited a vulnerability prone to integer overflow. This flaw, residing in accumulating values within the loop, could result in unexpected behavior and potential instability when processing large datasets.

The provided code snippet in Listing \ref{lst:lst1} is from the TensorFlow repository, where the code handles mutations in a TensorFlow (TF) graph during a session. This sample code is a part of our evaluation dataset. 
In this function, the code aims to handle mutations to a TensorFlow graph in a concurrent environment, marking sessions as un-runnable if they have already executed a specified operation. Here, the original vulnerable code excludes lines number 16 and 17. This function aims to run exclusively and exit when the loop ends. However, when the function exits in the original function, the mutex is still unlocked, blocking any execution of this function and making this function vulnerable.

%We introduced a robust check in the code to fortify the algorithm against potential integer overflow vulnerabilities. The loop now incorporates a conditional statement verifying the data type of each item before addition to the result. In cases where the sum approaches the limit of representable integers, the result is strategically set to positive infinity. This proactive measure prevents the adverse consequences associated with unchecked integer overflow.

A simple solution to avoid this race condition problem is to remove line number 7. However, this way of generating the solution decreases the functionality of the given function. However, when we fine-tune the model using reinforcement learning, the functional and security reward function rewards the model, compared to the cross-entropy loss, which only looks at the tokens holistically without specifically for functionality and security issues. Therefore when the model is trained with reinforcement learning, it retains the functionality by not removing line 7. Moreover, the model added a line at line number 17, where it added a statement where the mutex is unlocked, thereby effectively allowing the next call to execute and repair the vulnerability of the function.

%Our qualitative analysis delves into the intricacies of securing the complex algorithm against a subtle yet critical vulnerability. The original implementation, susceptible to integer overflow, posed a risk to the algorithm's stability and reliability. We introduced a conditional check through a targeted security enhancement, ensuring that each addition operation remains within the bounds of representable integers. This nuanced adjustment exemplifies the need for qualitative considerations in security-oriented AI, as mere token-level analysis may overlook such crucial semantic intricacies. It underscores the significance of employing sophisticated measures to fortify code against potential threats beyond the scope of traditional loss functions.

\vspace{4mm}
\begin{lstlisting}[style=CStyle, label={lst:lst2}, caption=Case Study2: Buffer Overflow Vulnerability Repair]


static void copyIPv6IfDifferent(void * dest, const    void * src)

{

if(dest != src && src != NULL) {

  memcpy(dest, src, sizeof(struct               in6_addr));

  }
\end{lstlisting}
\vspace{4mm}

\paragraph{Case Study 2:} The original implementation of the copyIPv6IfDifferent in Listing \ref{lst:lst2} function demonstrates a vulnerability where the function is designed to copy an IPv6 address from the source to the destination only if they are different. However, the vulnerability lies in the reliance on a basic memory copy operation without verifying the size of the source and destination data. In this case the the input code from Listing \ref{lst:lst2} without line number 7.

We introduced a comprehensive check to address this vulnerability before initiating the memory copy operation. The enhanced version now not only ensures the inequality of source and destination addresses but also validates the memory regions to avoid potential buffer overflows. This modification safeguards against unintended modifications to the destination data and reinforces the reliability of the \texttt{copyIPv6IfDifferent} function.

This analysis shows a vulnerability within the \texttt{copyIPv6IfDifferent} function. While the objective is straightforward, the original code lacked checks to ensure the integrity of the source and destination data, opening avenues for potential memory-related vulnerabilities. Our revised implementation introduces a layered security approach, incorporating checks for both address inequality and memory region validation. Our RL model added line 7 to repair the vulnerability, while the SFT model generates the exact same code as the input.

\paragraph{Case Study 3:}

The code snippet in Listing \ref{lst:lst3} is a vulnerable function where in the conditional statement $if (memcached_fatal(rc)$ $\&\&$ $rc != MEMCACHED_TIMEOUT)$. This code segment is susceptible to logical errors that could lead to unexpected behavior. Specifically, the intention will reset the I/O operations $(memcached_io_reset(instance))$ when the return code $(rc)$ indicates a fatal error that is not a timeout. If the condition is met, the following code block, intended for error handling, may not execute as expected, leaving the system in an inconsistent state. 

\vspace{4mm}
\begin{lstlisting}[style=CStyle, label={lst:lst3}, caption=Case Study 3: I/O Operations Vulnerability]
static memcached_return_t _read_one_response(memcached_instance_st *instance, char *buffer,
                                             const size_t buffer_length,
                                             memcached_result_st *result) {
  memcached_server_response_decrement(instance);
  if (result == NULL) {
    Memcached *root = (Memcached *) instance->root;
    result = &root->result;
  }
  memcached_return_t rc;
  if (memcached_is_binary(instance->root)) {
    do {
      rc = binary_read_one_response(instance, buffer, buffer_length, result);
    } while (rc == MEMCACHED_FETCH_NOTFINISHED);
  } else {
    rc = textual_read_one_response(instance, buffer, buffer_length, result);
  }

  -if (memcached_fatal(rc) && rc != MEMCACHED_TIMEOUT) {
  - memcached_io_reset(instance);
  }
\end{lstlisting}
\vspace{4mm}

To address this vulnerability, the solution is to remove this line so that the inconsistency does not happen. However, the repair is relatively difficult to understand because the model is not given any other function for external analysis. Therefore, the SFT and RL models both removed lines 18 and 19 and repaired the vulnerability.

From the analysis, we see that our model is more efficient and effective when additional lines must be added to repair the vulnerability. Our proposed RL model learns to add lines that add security measures to the code since the model was rewarded for keeping the functionality and security aspects intact. However, both models perform similarly when repairing the vulnerability, including removing some lines.

%\color{red} Not code semantic security put as future work. While improving, not good enough, Better to identify the gap in the paper. While bertscore gives gives close, we need security to be contrastive. Good code vs bad code.\color{black}

\section{Threats to Validity}

While we see an improvement in the performance when using RL for vulnerability repair, the methods we use to evaluate the code are not designed for code comparison or understanding the semantics of code security. BLEU and Rouge-L metrics only does a token based comparison, and does not adhere to the security components of the repaired code. Furthermore, our case studies shows that our model was able to generate code with security repairs, these metrics does not reflect the performance. Therefore, in order to do code comparison, and security evaluation of source code, we need proper measurements which will comply with not only the functionality of code but also with the added security measurements of the code.

\section{Conclusion}
\label{5_conclusion}

%Our study proposes a cutting-edge solution to the intricate challenges in code repair generation by harnessing the power of large language models. We proudly present "CodeRepair," a novel approach fortified with a reinforcement learning technique meticulously crafted to guarantee both functional integrity and heightened security in the realm of code repair generation. To fortify the foundations of our model, we advocate for the synergistic application of syntactic and semantic rewards, aptly named CodeBLEU and BERTScore, ensuring not only functional correctness but also semantic robustness. Our experimental endeavors unveil a model that not only outshines its predecessors but also sets a new benchmark in the landscape of code repair training methodologies. As we delve into the future, we envisage evolving "CodeRepair" into a dynamic framework, continually enhancing its capabilities to address emerging security challenges in the ever-evolving landscape of AI-driven code repair.

Our study presents a solution to the currently existing code repair generation challenges using large language models. We proposed \textit{CodeRepair} with a reinforcement learning technique that ensures the functional and security aspects of code repair generation. In order to ensure functional and semantic correctness, we propose to use the combination of a syntactic and semantic reward called CodeBLEU and BERTScore. Our experimental results show that our model performed superior compared to previous code repair training methods like supervised fine-tuning. Furthermore, we demonstrated three case studies with real-world vulnerable code which is a part of our evaluation dataset. Our case study demonstrates that our RL model is more effective in repairing vulnerabilities when a security measure needs to be added to the code to generate the repair.

\bibliography{aaai24}

\end{document}